# Minuet: A Scalable Distributed Multiversion B-Tree

Benjamin Sowell[*]
Cornell University
Ithaca, NY, USA
sowell@cs.cornell.edu

Wojciech Golab
Hewlett-Packard Labs
Palo Alto, CA, USA
wojciech.golab@hp.com

Mehul A. Shah[*]
Nou Data
Palo Alto, CA, USA
mashah@gmail.com

## ABSTRACT

Data management systems have traditionally been designed to support either long-running analytics queries or short-lived transactions, but an increasing number of applications need both. For example, online games, socio-mobile apps, and e-commerce sites need to not only maintain operational state, but also analyze that data quickly to make predictions and recommendations that improve user experience. In this paper, we present Minuet, a distributed, main-memory B-tree that supports both transactions and copy-on-write snapshots for in-situ analytics. Minuet uses main-memory storage to enable low-latency transactional operations as well as analytics queries without compromising transaction performance. In addition to supporting read-only analytics queries on snapshots, Minuet supports writable clones, so that users can create branching versions of the data. This feature can be quite useful, e.g. to support complex "what-if" analysis or to facilitate wide-area replication. Our experiments show that Minuet outperforms a commercial main-memory database in many ways. It scales to hundreds of cores and TBs of memory, and can process hundreds of thousands of B-tree operations per second while executing long-running scans.

## 1. INTRODUCTION

Modern applications are placing increased pressure on the time from data ingest to insight. They not only need to update operational state fast, but also need to quickly analyze that data for recommendations and predictions that improve user experience. For example, e-commerce sites not only track per-user shopping activity, but also mine that data to recommend related items for purchase. Socio-mobile applications track user location to recommend nearby businesses or predict traffic delays. Online multi-player games track player positions and interactions to price and sell virtual goods. Credit card companies need to catch fraud, online ad networks need to discover recent trends to target ads, and so on. For today's businesses, the ability to quickly gain insight and immediately optimize is increasingly a necessity rather than simply a competitive advantage. Moreover, the fresher the data, the more effective is the insight from the analysis.

The state-of-the-art in most deployments is to employ two drastically different data management systems for managing operational state and analytics. Organizations use transactional systems, and more recently, key-value stores, for short-lived operations on operational state [13], and they use data warehouses for long-running analyses on that data. This separation is largely a result of conflicting workload needs: random access for transactions versus scans for analytics. This separation also typically implies a lag of hours, if not days, from data ingest to insight.

To close this gap, we present Minuet, a main-memory, distributed B-tree that supports both transactions and in-situ analytics. Its salient features include:

- **Performance and scalability** (see Section 6): Minuet supports hundreds of thousands of transactions per second with latencies near 1ms. We show that its performance scales nearly linearly up to hundreds of cores and TBs of memory in a shared-nothing commodity cluster.
- **Copy-on-write snapshots** (see Section 4): Minuet offers consistent snapshots to enable "in-situ" analysis without compromising the performance of ongoing transactions. It mitigates the overhead of snapshot creation by sharing snapshots intelligently between concurrent queries while still ensuring strict serializability. To reduce overheads even further, users can choose to execute queries using slightly stale snapshots.
- **Writable clones** (see Section 5): The snapshots themselves are writable, so users can create branching versions of data. Like revision control but for B-trees, this feature has many uses. Analysts can use it for "what-if" analysis in forensics or forecasting, e.g. what happens if I rebalance my investments? System developers can use it for wide-area replication and sharing, and archiving.

Minuet can offer this combination of features because it keeps its data entirely in memory. Unlike disks, memory can simultaneously provide low-latency random access and fast sequential access. Moreover, this approach is practical today given the availability of large main memories; most operational data sets can fit in the memory of a cluster [38]. Although others have also proposed such hybrid memory-based systems (e.g., [25]), Minuet is the first to scale across a cluster of servers and offer writable, branching clones.

Minuet is based on a prior scalable distributed B-tree [5] built using the Sinfonia data sharing service (Section 2) [6]. In this paper we extend this B-tree with a novel concurrency control optimization that enables scalable insertions and updates. Furthermore, we add support for long scan operations using copy-on-write snapshots that can optionally be made strictly serializable. Our experiments (Section 6) show that Minuet easily outperforms a modern commercial main-memory database. Minuet can execute hundreds of thousands of low-latency B-tree operations per second and scales linearly to hundreds of cores. By leveraging snapshots, it can also execute long-running scans concurrently with short update transactions with limited interference. This combination of features is a big step towards unifying operational and analytics systems, thereby allowing organizations to make faster and more insightful data-driven decisions.

## 2. OVERVIEW AND BACKGROUND

In this section, we describe the architecture of Minuet and review its building blocks. Minuet organizes data in a scalable distributed multi-version B-tree, and exposes a key-value interface with support for range queries. The system architecture is shown in Figure 1, and consists of *clients*, *proxies*, and *memnodes*. Clients issue requests for transactional B-tree operations. Proxies execute these operations on behalf of clients by accessing B-tree state stored at memnodes. Our prototype uses at its core the Sinfonia platform [6], which comprises the memnodes and a library that proxies use to communicate with memnodes. Sinfonia provides a lightweight transactional interface, indicated in Figure 1 by the dashed line. Proxies use these lightweight transactions to implement more powerful transactions, and then use the latter to execute B-tree operations—an approach based upon [5]. Components of the system communicate using remote procedure calls and may be physically co-located or separated by a data center LAN.

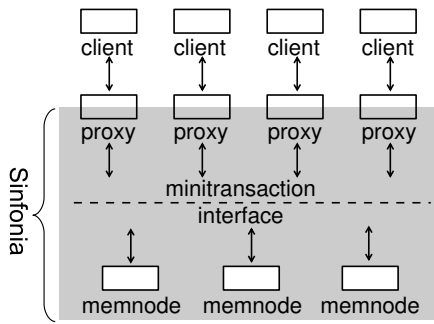

**Figure 1: Minuet architecture.**

Minuet's architecture and design are based upon a prior distributed B-tree [5]. In the remainder of this section, we give a detailed overview of this prior system, including the Sinfonia platform, a dynamic transaction layer, and transactional B-tree algorithms. Later on in Sections 3–5, we describe our own contributions: an extension of the dynamic transaction layer that enables greater concurrency, and B-tree algorithms that exploit this extension for better scalability as well as for supporting multi-versioning.

### 2.1 Sinfonia

Sinfonia is a distributed data sharing service comprising a set of storage nodes called *memnodes*, and an application library for executing data operations [6]. Each memnode exports an unstructured byte-addressable storage space similar in spirit to an array of bytes. The Sinfonia library provides fault-tolerant transactional access to this address space and is linked into the proxies. The Sinfonia system provides fault tolerance by masking network failures and memnode crashes, which simplifies the development of robust distributed applications on top of the platform.

Sinfonia applications act on the shared state stored at memnodes by executing *minitransactions*. A minitransaction can read, compare, and conditionally update data at multiple memory locations, possibly at multiple memnodes running on different servers. The application specifies the addresses of the memory locations ahead of time, and the updates are applied atomically only if all the comparisons evaluate positively (or there are no comparisons). In Minuet, a proxy might use a minitransaction to read a B-tree node from a memnode during a B-tree traversal, as well as to update one or more B-tree nodes in the course of a B-tree update or insertion operation.

Sinfonia uses a two-phase protocol to execute and commit distributed minitransactions. In phase one, memnodes lock memory regions touched by the minitransaction and evaluate comparisons. Sinfonia aborts the minitransaction immediately if a lock is busy or a comparison fails. In the former case, the application library retries the minitransaction automatically and transparently to the application. In the latter case, the library passes control back to the application and indicates which comparisons failed. The two-phase protocol is collapsed down to a single phase automatically whenever only one memnode is involved.

### 2.2 Dynamic Transaction Layer

As hinted earlier, a minitransaction can be used to execute part of a B-tree operation such as fetching a B-tree node from a memnode. However, since the memory locations touched by a minitransaction must be specified in advance, a single minitransaction cannot traverse a B-tree. Aguilera et al. [5] describe how to use minitransactions to construct a more powerful *dynamic transaction* that can read and write objects (e.g., B-tree nodes) arbitrarily using optimistic concurrency control with backward validation [20]. Each dynamic transaction maintains a read set and write set of objects it touches. A transactional read first tries to read the object locally from the write set or read set, and on failure it triggers a minitransaction that fetches that object from a memnode and adds it to the read set. Transactional writes place objects in the write set and defer updating the memnodes until the entire dynamic transaction is committed. Committing entails executing a minitransaction that (1) validates the read set (i.e., verifies that the objects in the read set are identical to their "master copies" at memnodes); and (2) if the validation succeeds, copies the objects in the write set to memnodes.

Validating the entire read set atomically with step (2) ensures that dynamic transactions are serializable. In order to reduce the CPU and network overhead due to validation, objects can be tagged with sequence numbers that increase monotonically on update, and comparisons are based solely on these sequence numbers. Furthermore, it is possible to piggy-back validation onto minitransactions triggered by transactional reads, in which case step (1) can be skipped entirely on commit whenever the write set is empty.

Dynamic transactions can be used to transform any centralized data structure implementation into one that is distributed among multiple memnodes and can be accessed by many clients/proxies in parallel. Doing so specifically on top of Sinfonia automatically provides fault-tolerance and enables scalable throughput for read-only transactions. (For other types of transactions, scalability is possible but not automatic, as in any optimistic concurrency control scheme.)

### 2.3 Distributed B-Tree

Aguilera et al. describe a scalable distributed B-tree implemented using their dynamic transaction layer [5]. Their implementation distributes the B-tree by placing different B-tree nodes at different servers. A distributed memory allocator decides the placement of B-tree nodes in a way that balances load. The allocator itself is a data structure implemented using dynamic transactions, and can be shared easily by multiple applications.

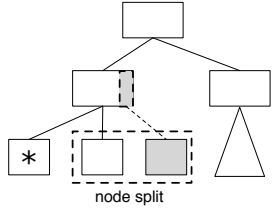 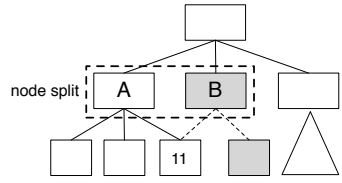 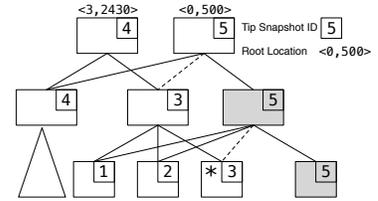

Figure 2: Unnecessary abort example.   Figure 3: Incorrect traversal example.   Figure 4: Copy-on-write example.

Fundamental B-tree operations (get, put, remove) are obtained by wrapping dynamic transactions around single-threaded code. This entails using transactional reads and writes to fetch and update B-tree nodes, and introducing safety checks to ensure that the code does not crash or hang when transactional reads return inconsistent data, which may happen during a transaction that will eventually abort. Aside from that, the main implementation challenge lies in optimizing for performance by reducing the number of network round trips needed to commit a B-tree operation, and by avoiding contention.

For lookup and update operations, the B-tree of Aguilera et al. scales nearly linearly to hundreds of servers thanks to two specific optimizations. First, to reduce communication, they cache internal B-tree nodes (i.e., non-leaf nodes) at the proxies[1]. The cache is part of the proxy application code, and does not ensure coherency across proxies or across objects cached at the same proxy. Second, to speed up the validation step during dynamic transaction commitment, they replicate the sequence number of each internal B-tree node at each memnode. This ensures that the server storing the root node does not become a bottleneck. Furthermore, it reduces the "minitransaction spread" by allowing a B-tree operation to commit at a single memnode, namely the one storing the leaf node, unless a B-tree node splits. This enables one-phase commitment, which not only reduces the number of network delays but also shortens dramatically the period of time for which memnodes must hold locks on B-tree nodes.

Lazy caching of B-tree nodes at proxies and eager replication of sequence numbers at memnodes ensure that most B-tree operations can be committed using only one or two round trips to a single server. In the best case, a lookup operation traverses the B-tree in-cache up to leaf level, then fetches the leaf and validates the path traversed in the same minitransaction. Similarly, an insertion fetches the leaf in one round trip, and then commits the updated leaf in one additional round trip provided that the leaf does not split. Optimistic concurrency control works particularly well in this context because the interior levels of the B-tree change infrequently.

## 3. TRANSACTIONS WITH DIRTY READS

The concurrency control mechanism used in the B-tree of Aguilera et al. [5] suffers from two important drawbacks. First, replicating the sequence numbers of all internal B-tree nodes at each memnode makes any operation that updates a sequence number, such as a node split or merge, expensive to commit. Since such operations engage all memnodes, they do not scale and are especially susceptible to contention. They also cannot make progress at all if even a single memnode becomes unavailable.

---
[1]Note that in their implementation, they did not separate the client and proxy, and ran all application code, including caching, at the client. We will continue to use client to refer only to the application or process issuing requests.

Second, the concurrency control scheme is conservative, and may abort transactions unnecessarily. Consider Figure 2, and suppose that one transaction reads the node marked "*" by traversing the tree from the root. If another transaction concurrently causes a sibling to split (indicated in the figure using dashed lines), then the original transaction must abort, since the parent has been updated with the new child and thus the first transaction's copy will no longer validate. Note that the abort occurs even though the traversal reached the correct leaf node, which is not modified by any concurrent transaction. Furthermore, one operation can cause the other to stall when their underlying minitransactions contend for a lock on the parent B-tree node. As explained earlier, a minitransaction that encounters a busy lock is re-executed, incurring at least two additional network delays. Such stalls delay the commitment of B-tree operations, which tends to increase the likelihood of an abort.

To address these challenges, we introduce *dirty reads* to the set of transactional operations supported by dynamic transactions (which in [5] comprised only reads and writes). A dirty read allows a proxy to fetch an object from its local cache or from a memnode without adding the object to its read set for validation. (If the object is written later on, it will first be added to the read set.) With dirty reads, dynamic transactions may no longer be serializable, but we observe that it is possible to make B-tree operations strictly serializable even if dynamic transactions provide weaker isolation.

To execute B-tree operations, proxies traverse the tree from the root node down to one level above the leaves using dirty reads, and then read the leaf node using an ordinary transactional read. Thus, in the common case when the transaction does not cause a B-tree split, the read set contains only a single leaf node rather than the entire path from root to leaf. This greatly reduces the number of objects that must be validated, in most cases down to only one object—the leaf node. Moreover, we no longer need to replicate the sequence numbers of internal B-tree nodes. This makes insertion operations even faster and more resilient against failures, and also reduces the space overhead due to the replicated sequence number tables. (The table size per server is proportional to the aggregate capacity of the system, and so as the number of servers increases, the table size eventually outgrows the capacity of one server.)

Traversing a B-tree using dirty reads requires additional safety checks to deal with inconsistent data because inconsistencies no longer force transactions to abort. For example, a traversal searching for a given key $k$ may visit B-tree nodes that do not lie on the path from the root to the leaf node responsible for key $k$, and then commit successfully. Although this is benign in some cases, it can be fatal in others, such as when the traversal ends at the wrong leaf node, or a leaf splits and the traversal path does not include the correct parent node. For instance, consider the scenario depicted in Figure 3. In this example, one transaction searches for key 11, and another inserts a new key which causes an internal node split. It is possible for these transactions to be interleaved in such a way that the first transaction reads node A after the split and incorrectly concludes that key 11 is not present in the tree. If we use a dirty read of

**Function** Traverse($R$, $k$, $T$)

| | |
|---|---|
| **Input:** | $R$ – pointer to root of B-tree with at least two levels |
| | $k$ – search key |
| | $T$ – dynamic transaction |
| **Output:** | sequence of B-tree nodes traversed from root to the leaf responsible for key $k$ in B-tree rooted at $R$, or else $\perp$ if $T$ aborted |

```
 1  curPtr := R
 2  curNode := T.DirtyRead(internal node at curPtr)
 3  ret := ⟨curNode⟩
 4  while isInternal(curNode) do
 5      if k < lowFence(curNode) OR k > highFence(curNode) then
 6          T.Abort(), return ⊥
 7      else
 8          nextPtr := child of curNode responsible for key k
 9          if height(curNode) > 1 then
10              nextNode := T.DirtyRead(internal node at nextPtr)
11          else
12              nextNode := T.Read(leaf node at nextPtr)
13          end
14          ret := ret ∘ ⟨nextNode⟩
15          if height(nextNode) ≠ height(curNode) − 1 then
                // Fatal inconsistency!
16              T.Abort(), return ⊥
17          end
18          curNode := nextNode
19          curPtr := nextPtr
20      end
21  end
    // Reached leaf node
22  if k < lowFence(cur) OR k > highFence(cur) then
23      T.Abort(), return ⊥
24  end
25  return ret
```

**Figure 5: Transactional B-tree traversal using dirty reads.**

node A, then this transaction might commit with an incorrect result.

In order to prevent such anomalies, we introduce *fence keys* at each node in the tree [28, 17]. Fence keys define the range of keys that a node is responsible for, whether or not they are present in the tree. During a search, we compare the fence keys at a node with the search key $k$, and abort the transaction if $k$ is out of range. This guarantees that we either reach the correct leaf node or abort, and hence prevents anomalies like the one shown in Figure 3.

Our B-tree traversal algorithm is presented in detail in Figure 5. In this algorithm, $T$ denotes a dynamic transaction, and the operations $T$.Read and $T$.DirtyRead denote a transactional read and dirty read, respectively.

## 4. SNAPSHOTS

Serializable transactions, such as those provided by Minuet, simplify greatly the development of complex distributed applications. However, they are also inherently expensive and can lead to poor performance, especially for workloads that include long-running transactions. As an increasing number of organizations recognize the importance of real-time data analysis, this kind of mixed workload will become increasingly common.

Similarly to several recent systems [25, 10], Minuet addresses this challenge by executing long-running queries, such as scans over indexes, against consistent snapshots of the data. Proxies can create snapshots on demand in such a way that queries always appear to act on the latest data, which guarantees strict serializability [34]. This means that transactions (including queries) not only appear to execute in some serial order (as per serializability), but moreover this order is consistent with the "happens before" relation over transactions (i.e., if $T_1$ ends before $T_2$ begins then $T_2$ does not appear to take effect before $T_1$).

Snapshots provide a consistent view of the B-tree at a fixed point in time, and can be used for a variety of applications, including archival and WAN replication. Most importantly, we can use snapshots to isolate analytics queries from the ambient OLTP workload. In this case, we can mitigate the cost of snapshot creation by sharing snapshots (see Section 4.3), or by deliberately executing queries against existing slightly stale snapshots (see Section 6.3).

In the remainder of this section, we describe how to create read-only snapshots in Minuet in a strictly linearizable way. We then discuss the extensions necessary to support writable snapshots (i.e., clones) and branching versions in Section 5.

### 4.1 Copy-on-Write Snapshots

Since Minuet stores data in a distributed B-tree, we can leverage existing copy-on-write checkpointing techniques to take consistent snapshots efficiently [36]. At a high level, when a new snapshot is created, every B-tree node is subsequently copied before being updated so that the snapshot is not overwritten.

A snapshot is a version of the B-tree identified with a 64-bit *snapshot id*, which indicates the (total) order in which snapshots are created. The latest or *tip snapshot* is writable, and all prior snapshots are read-only. Creating a new snapshot in this context means making the tip snapshot read-only and creating a new tip with id one higher than the previous tip. The new tip shares each B-tree node with the previous snapshot until the B-tree node is overwritten in the tip snapshot. Each B-tree node is annotated with the snapshot id at which it was created, which happens by way of a B-tree split or a copy-on-write. To keep track of the tip snapshot, we store its id and the location of the corresponding root node at a well-defined location in Sinfonia's address space.

Reads can be performed against any snapshot, though the reader is responsible for keeping track of the location of the root node for read-only snapshots. Strictly serializable or *up-to-date* reads must act on the tip snapshot. In this case, the read fails if a new tip is created concurrently. When describing operations on a specific B-tree node, we will say that a read or write occurs *at* a snapshot $s$ to make it clear on which snapshot we are operating. For an up-to-date operation, $s$ will be the tip snapshot.

When we update a B-tree node at snapshot $s$, we first compare $s$ with the snapshot id stored in the node. If $s$ is larger than this value, we copy the node and update the copy, tagging the new node with snapshot id $s$. We then adjust the parent of the old node so that it points to the new node. This update is also performed at snapshot $s$, which may force another copy. In general, an update to a leaf node may require copying all nodes on the path from the root to the leaf, though the root itself will never be copied as it is copied already during snapshot creation (described below).

Figure 4 illustrates the copy-on-write procedure. In this diagram, the root of each snapshot is annotated with its location, given as a pair containing the memnode and offset. Each node is shown with its snapshot id, and the leaf node marked with a "*" is being updated at snapshot 5. Since the leaf node was created at snapshot 3, it must be copied to a new node shown in gray. Similarly the parent node must be copied before its child pointer can be updated. The child pointer in the root must also be updated, but since the root is already at snapshot 5, it does not need to be copied. The old child pointers are shown as dashed lines.

This copy-on-write procedure can be implemented in Minuet using dynamic transactions. In order to perform an up-to-date read or write, a proxy adds its cached copy of the tip snapshot and corresponding root location to the transaction's read set. Thus, if another transaction creates a new tip concurrently, the validation will fail, and the transaction will be aborted and retried. In contrast, when a proxy reads from a read-only snapshot (i.e., a snapshot earlier than the tip), the transaction can only abort if the proxy's cache con-

**Function** CreateSnapshot(*sid, loc, T*)

| | |
|---|---|
| **Input:** | $T$ – dynamic transaction |
| **Output:** | $sid$ – snapshot id of the snapshot created |
| | $loc$ – location of root node for the snapshot created |

1  T.Read($tipSnapshotID$)
2  T.Read($tipSnapshotRootLoc$)
3  $sid := tipSnapshotID$
4  $loc := tipSnapshotRootLoc$
5  $tipSnapshotID\mathrel{+}=1$
6  $newRootLoc :=$ Allocate($NodeSize, T$)
7  CopyRoot($tipSnapshotID, tipSnapshotRootLoc, newRootLoc, T$)
8  $tipSnapshotRootLoc := newRootLoc$
9  T.Write($tipSnapshotID$)
10 T.Write($tipSnapshotRootLoc$)
11 **return** ($sid, loc$)

**Figure 6: Snapshot creation algorithm.**

tained a stale copy of a B-tree node corresponding to that snapshot. This can happen if one proxy cached an inner B-tree node before another proxy modified that node and then took a snapshot.

Creating a snapshot simply requires incrementing the tip snapshot id. The value before incrementing will be the id of the read-only snapshot, and the value one greater will be the new tip snapshot id. Pseudocode for this procedure is shown in Figure 6. The objects $tipSnapshotID$ and $tipSnapshotRootLoc$ store the tip snapshot id and root location, respectively. Lines 1 to 4 read the tip snapshot id and assign the output variables, while line 5 increments the tip snapshot id. Additionally, during snapshot creation, we copy the root of the tree and update the root location of the tip snapshot (lines 6 to 8). This could be deferred until the first update, but copying the root at snapshot creation time ensures that the root node for the tip snapshot remains at a fixed position in Sinfonia's address space. This property simplifies application code, including the snapshot borrowing technique described later on. Note that in Figure 6, the dynamic transaction used to access the tip snapshot id is a parameter of the CreateSnapshot procedure—the snapshot is not actually created until $T$ is committed successfully.

Note that every write and all up-to-date reads must validate the tip snapshot id and root location. To avoid a contention hotspot and to ensure that most B-tree operations can commit at a single server, we replicate these objects across all memnodes just as Aguilera et al. replicate the sequence numbers for internal B-tree nodes [5]. Abusing notation slightly, in Figure 6 we denote transactional reads and writes on these replicated objects by T.Read and T.Write (as for ordinary objects), with the understanding that the reads can access any replica and the writes must update all replicas.

Replicating the tip snapshot id and root location increases the cost to update the tip snapshot id, as we must write to all memnodes atomically—a contention-prone operation. However, we expect the frequency of snapshot creation operations to be much lower than the frequency of B-tree gets and puts, and so we expect the benefit of efficient validation to exceed the additional update cost. Furthermore, we mitigate the cost by updating the replicated snapshot id using a special *blocking minitransaction*, which waits at the memnode for locks to be released instead of aborting when there is contention for a lock. The waiting time is bounded by a threshold small enough so that blocking minitransactions do not trigger Sinfonia's recovery mechanism unnecessarily. On occasion the threshold is exceeded, in which case the blocking minitransaction simply aborts, just like an ordinary minitransaction would.

### 4.2 Dirty Traversals and Snapshots

One of the ways in which Minuet differs from previous work is its use of dirty traversals to achieve higher concurrency. However, we must take care when implementing this technique with snapshots in order to avoid additional anomalies. Previously, we addressed the case in which a traversal ended up at the wrong leaf node due to a node split, but now we must also consider the scenario in which a traversal ends up at the correct leaf node in the wrong snapshot.

For example, in Figure 4, if the copy-on-write is performed concurrently with a search terminating at node marked "*", it is possible that the search would see the old snapshot of the internal node rather than the new tip snapshot. If the copy-on-write commits before the search reaches the leaf, then it is possible that the leaf node will validate successfully, even though it has an incorrect snapshot id. Fence keys are not sufficient to solve this problem, as the leaf node may cover the correct key range but in a stale snapshot.

To address this problem, we introduce one additional piece of state in each B-tree node: the snapshot id for which the node has been copied, if any. For instance, in Figure 4, we would store 5 in both of the nodes that were copied (in addition to the snapshot id for which each node was created), as their copies have snapshot id 5. Note that this quantity is well defined, as each node can be copied at most once.

During a read or write at snapshot $s$, if the search encounters a node that has been copied to a snapshot id less than or equal to $s$, then it aborts because the traversal should visit the copy (or a copy of the copy, etc.) instead. Otherwise, if the node has the appropriate fence keys, then it is guaranteed to be on the correct traversal path.

Note that while this procedure guarantees correctness, it can lead to performance problems when validation is done using the technique described so far. In particular, now when a node is copied due to a copy-on-write operation, we must change its sequence number since we have updated its state. This could cause operations (e.g., scans over multiple leaf nodes) on old snapshots to fail to validate and abort unnecessarily. Fortunately, since such snapshots are read-only and since we always fetch leaf nodes directly from Sinfonia, we can avoid validating leaf nodes entirely by using appropriate safety checks on fence keys.

### 4.3 Borrowed Snapshots

While we expect the frequency of snapshot creation operations in Minuet to be low relative to gets and puts, such operations are quite heavyweight, as they involve updating the replicated snapshot id and root location atomically at all memnodes. Thus, relatively few concurrent snapshot creation requests can greatly degrade the performance of both the snapshots and get/put operations, which must validate the tip snapshot id. To address this potential bottleneck, we apply two optimizations. First, we create all snapshots one-at-a-time using a centralized service, which reduces dramatically contention on the replicated tip snapshot id object. Second, we observe that since the snapshots returned by this service are all read-only, several operations can share the same snapshot without interfering. While other systems allow users to explicitly request that multiple queries be run against the same snapshot (e.g., query sessions in HyPer [25]), we introduce a technique that automatically decides which queries should share a snapshot. The technique, called *borrowed snapshots*, enables one query to use a snapshot created by a concurrent query provided that this preserves strict serializability.

The high-level idea behind borrowing is as follows. Suppose that clients $A$ and $B$ try to create snapshots concurrently, and $B$'s request is queued behind $A$'s. If $A$'s snapshot is created while $B$ is already waiting in the queue, then $A$'s snapshot reflects the state of affairs at some point in time during the execution of $B$'s request. Consequently, $B$ can borrow this snapshot (instead of creating its own) safely without compromising strict serializability.

The pseudocode for creating a snapshot with borrowing is shown

**Function** CreateSnapshotProc()

| | |
|---|---|
| **Output:** | $sid$ – the snapshot id that was created or borrowed |
| | $loc$ – Sinfonia address of the root node for $sid$ |
| **Variables:** | $sid, loc$ – snapshot id and root node address, shared/static |
| | (initially $sid = 0$ and $loc$ is the address of the initial root node) |
| | $mutex$ – mutual exclusion lock, shared/static |
| | $numSnapshots$ – atomic integer, shared/static (initially 0) |

1  $tmpNum_1 := numSnapshots$
2  lock($mutex$)
3  $tmpNum_2 := numSnapshots$
4  **if** $tmpNum_2 < tmpNum_1 + 2$ **then**
    // unable to borrow, must create new snapshot
5     Dynamic transaction $T$
6     CreateSnapshot($sid, loc, T$) // see Figure 6
7     **if** $T$.Commit() = false **then** continue at line 5
8     ++$numSnapshots$
9  **else**
    // safe to borrow last snapshot, nothing to do
10 **end**
11 $retSid := sid$
12 $retLoc := loc$
13 unlock($mutex$)
14 **return** $retSid, retLoc$

**Figure 7: Snapshot creation service (SCS).**

in Figure 7. CreateSnapshotProc is a remote procedure exported by one or more servers in the cluster. Multiple servers can be used for fault tolerance, but in order to avoid contention (on the replicated tip snapshot id object) all proxies should route snapshot requests to the same server, for example one chosen using distributed leader election. Proxies invoke CreateSnapshotProc concurrently, and portions of the procedure are executed in parallel by the server's executor threads. The main portion, however, is executed inside a critical section between lines 2 and 13. Inside the critical section, the executor thread decides at line 4 whether it will create a new snapshot or borrow (i.e., reuse) the snapshot created most recently. The decision depends on the value of the atomic integer $numSnapshots$, which records the total number of snapshots that have been created by all threads. A thread reads this counter first at line 1, before the critical section, and then again at line 3, inside the critical section. If the counter increases by two or more between these two reads, then it follows that in the mean time some other thread has started and finished a call to CreateSnapshotProc at line 6. Thus, the former thread can borrow the snapshot created by the latter thread. Indeed this occurs at lines 11–12. Otherwise, the executor thread calls CreateSnapshotProc directly at line 6 and returns the output of that call.

Note that the decision to share a snapshot among two transactions can be made both inside the snapshot creation service (SCS), as described above, and also in a distributed fashion at the proxies. For example, if transactions $T_1, T_2$ try to create snapshots concurrently at proxy $A$, and in parallel transactions $T_3, T_4$ do so at proxy $B$, then $A$ and $B$ could allow only $T_1$ and $T_3$ to invoke the SCS, $T_3$ could borrow from $T_1$ inside the service, and finally $T_2$ and $T_4$ could borrow from $T_1$ and $T_3$ (respectively) at the proxies. For simplicity, in this paper we consider sharing only at the SCS and not at the proxies. We evaluate the performance of this scheme empirically in Section 6.

### 4.4 Garbage Collection

As the system runs and snapshots are created, Minuet will eventually run out of memory. To address this, we periodically garbage collect old snapshots. Minuet records a global *lowest snapshot id*, which is the smallest snapshot id to which a client can issue queries. This value can be set by the user, or set to increment automatically, for instance always supporting queries over the ten most recent snapshots. A background process periodically goes through the B-tree nodes stored at each memnode and collects those that have been copied to a snapshot less than or equal to the lowest snapshot id. These nodes are safe to delete, as they are never referenced by a snapshot newer than the lowest snapshot id. Any other B-tree node may be referenced from a snapshot higher than the lowest snapshot id, and for that reason cannot be deleted.

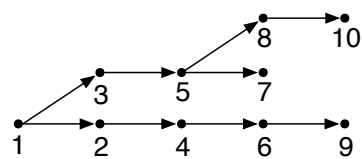

**Figure 8: Example version tree.**

## 5. WRITABLE CLONES

Creating sequential read-only snapshots is useful for many analytics tasks, but for more complex problems, it is often desirable to modify parallel versions of the data directly. Many users are familiar with the branching functionality found in most version control systems, and this feature can be useful for more sophisticated analytics tasks as well. For example, an analyst working on a predictive model might wish to validate a hypothesis by experimenting with slightly modified data. While she could export the data and modify it separately, there are several advantages to creating a branch within the same system instead. First, if we change only a small fraction of the data, it may be much more efficient, in both time and space, to use a copy-on-write approach rather than exporting the entire data set. Maintaining several versions in the same system also allows us to issue transactional queries across different versions of the data, which may be useful for integrity checks and to compare the results of an analysis.

In this section we describe how to add support for branching versions to Minuet using a slight modification of the copy-on-write method discussed in the previous section. This type of branching has been explored before in the context of single-machine databases [26, 23], but to our knowledge has never been incorporated into a distributed data store like Minuet.

### 5.1 Branching Versions

We support branches using the same basic copy-on-write algorithm described in Section 4. The main difference is that in addition to creating snapshots with monotonically increasing ids, it is also possible to create a branch from an existing read-only snapshot. This leads to a (logical) tree of (physical) B-tree versions, as illustrated in Figure 8. The internal vertices of this version tree correspond to read-only snapshots, while the leaves are writable or tip snapshots of the B-tree. Thus, in Figure 8, clients can write to snapshots 7, 9, and 10, while the other versions are read-only. Clients can create a new branch from any existing version, which will create a new leaf in the version tree to which updates can be applied.

Since there is no longer a natural total order on snapshots, it is no longer obvious how to assign ids to snapshots. One option is to encode the structure of the version tree into the snapshot id, for example using the Dewey id labeling scheme common in XML processing. Unfortunately, this would mean we could no longer bound the size of a snapshot id, as it would grow with the number of branches. To avoid this, we will continue to impose a total order on the snapshots by serializing snapshot creation and assigning snapshot ids using a monotonically increasing integer. (A 64-bit integer is large enough to preclude overflow in practice.) This limits somewhat the concurrency of snapshot creation, but we expect this to be a reasonable trade-off as snapshots are relatively infrequent.

Since there is no longer a unique tip snapshot id to validate, we must also change how we perform up-to-date reads and writes. For instance, if a client has read data at snapshot 5 in the version tree from Figure 8, the "correct" tip snapshot might be either 7 or 10. By default, we follow the branch created first (i.e. the one with the lowest snapshot id) when retrying a dynamic transaction, but the user can explicitly specify an alternative. This defines a *mainline* of snapshot ids in the version tree that starts at the root and ends at the tip snapshot used for up-to-date operations.

To keep track of the version tree, we introduce a *snapshot catalog* containing meta-data about each snapshot. This catalog includes the id of each snapshot, the location of the root node, and the first branch created from the snapshot (if any). We call this last value the *branch id* of a snapshot. When the branch id is NULL, no branches have been created from the snapshot, and it is writable.

We store the catalog in Sinfonia using a separate B-tree without snapshot support. This automatically ensures that the catalog is durable and available. As with the tip snapshot id in Section 4, popular snapshot ids in the catalog may become bottlenecks during validation. For this reason, we modify the tree that stores the catalog to replicate the leaf nodes across all memnodes and cache them at proxies. This improves performance, since we expect the leaves to be updated infrequently but read frequently. We do not replicate internal nodes because we use dirty traversals, as described in Section 3, to limit the number of internal validations.

In our B-tree with branching versions, basic B-tree operations are similar to those described in Section 4. Up-to-date reads and writes must validate the branch id of the desired snapshot and retry if the validation fails (i.e., if the branch id becomes non-null). As before, B-tree nodes are copied before they are updated, and we assign the copy the snapshot id at which the write occurs.

To create a new snapshot, we increment the global snapshot id and create a new entry with this id in the catalog. We must also allocate a root node to anchor the new snapshot, and update the branch id of the snapshot from which it was created. These operations must be performed atomically using a dynamic transaction. Note that creating a new branch is identical to creating a new snapshot. Intuitively, creating a new snapshot simply creates the first branch from an existing snapshot, and additional branches are created in the same way, except that the branch id does not change.

The technique of borrowed snapshots from Section 4.3 can be used equally well with branching versions. In fact, the catalog simplifies matters somewhat because borrowing can occur at each of the tip snapshots independently. Note, however, that borrowing cannot be used when creating a new branch, as branching creates a new writable snapshot that cannot be shared.

Unfortunately, the dirty traversal technique described in Section 4 does not work out-of-the-box in the context of branching versions. In the remainder of this section, we explain the technical issues that arise, as well as a suitable workaround.

## 5.2 Dirty Traversals

Recall that in Section 4, every B-tree node is tagged with both the snapshot id at which the node was created, and the minimum snapshot id to which the node was copied. Consider a traversal at snapshot $t$ that visits a B-tree node that was created at snapshot $x$ and later copied to snapshot $y > x$. In Section 4, the traversal continues as long as $x \leq t < y$ and the search key is within the range defined by the fence keys at the B-tree node, otherwise it aborts. For example, if $t \geq y$ then the traversal should instead be at a B-tree node created at version $y$ (or later). With branching versions, the same rule does not apply because snapshot $y$ may have been created in a branch different from the one containing snapshot $t$, such as in Figure 8 with $x = 1$, $y = 5$ and $t = 9$.

The naive analog of dirty traversals for branching versions would record for each B-tree node the snapshot id $x$ at which the node was created, and the *set* of snapshot ids (descendants of $x$ in the version tree) to which the node was copied. We will refer to this set as the *descendant set* of a B-tree node. Given this set and a method to determine whether two snapshot ids lie along the same directed path in the version tree (i.e., one derives directly or indirectly from the other), it is possible to decide correctly whether a B-tree traversal should continue or abort immediately at some tree node. (As usual, we also take fence keys into account.) Unfortunately the descendant set can grow without bound as snapshots are created and the version tree expands, which makes the naive technique impractical.

Our approach to making dirty traversals work with branching versions is to modify the above naive approach in a way that bounds the size of the descendant set. Part of the solution is to restrict the branching factor of the version tree, say to some constant $\beta$. However, this alone does not bound the size of the descendant set, which could be as large as the number of tip snapshots in the version tree. For example, in Figure 8 the version tree has branching factor $\beta = 2$ and yet if a B-tree node created at snapshot id 1 is copied-on-write to snapshots 7, 9 and 10, then the descendant set at that B-tree node would contain $3 > \beta$ entries.

The second part of our solution is to enforce the following invariant:

> If a B-tree node created at snapshot $x$ is copied to some subset $C$ of descendants of $x$ in the version tree, then there is a subset $C' \subseteq C$ of size at most $\beta$ such that for every $y \in C$, $C'$ contains an ancestor of $y$.
> (A vertex in the version tree is its own ancestor/descendant.)

We enforce this invariant by introducing *discretionary copy-on-write operations*. Going back to the example of Figure 8, if a B-tree node is created at snapshot $x = 1$, and then copied-on-write at 7, 9 and 10, then the invariant with $\beta = 2$ requires that the node also be copied at a common ancestor of 7 and 10 that is different from 1 (i.e., either 3 or 5). For example, one could have $C = \{3, 7, 9, 10\}$ and $C' = \{3, 9\}$, with the descendant set of the B-tree node equal to $C'$. In practical terms, this means that if the B-tree node is overwritten first at snapshot 7 and later at 10, then the latter write triggers a discretionary copy-on-write at, say, snapshot 3, and updates the descendant sets of the copies at both 1 and 3. The end result is as if the B-tree node created at snapshot 1 was first copied at snapshot 3, and subsequently at 7 and 10.

Our adaptation of dirty traversals to branching versions incurs space overhead in two ways. First, it requires additional space at each B-tree node to record the descendant set, which consists of up to $\beta$ snapshot ids. Second, it allocates additional B-tree nodes during discretionary copy-on-write operations. The latter increases the space overhead of copy-on-write by at most a factor of two because at most one discretionary copy is made for a given B-tree node for each ordinary copy. In cases where a user creates a branch in the version tree temporarily and then deletes it, the discretionary copies can be garbage collected to save space.

Finally, we point out that the size of the descendant set can be controlled in a fine-grained manner according to a user-defined policy. For example, when a user creates a side branch temporarily for "what-if" analysis, a larger $\beta$ is helpful because it simplifies creation of sub-branches and reduces the frequency of discretionary copy-on-writes. On the other hand, using a large $\beta$ along the mainline of the version tree (e.g., $\langle 1, 2, 4, 6, 9\rangle$ in Figure 8) might be overkill because for most of the B-tree nodes at those versions, the descendant sets might contain only one or two elements. Thus, for snapshots along the mainline it may make sense to fix $\beta = 2$.

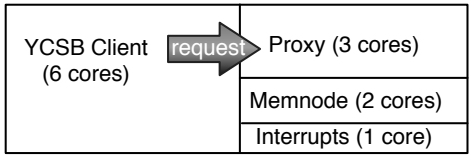

**Figure 9: Minuet experimental server utilization.**

## 6. EXPERIMENTS

In this section we evaluate the performance of Minuet on a variety of synthetic microbenchmarks.

### 6.1 Experimental Setup

We performed all experiments on a cluster of commodity machines. Each machine has two six-core 2.67 GHz Intel Xeon X5650 processors, 96 GB of RAM, and a 10 GigE network interface. Each core has two hyperthreads, but in order to limit contention, we pin processes so that only one hyperthread per core is in use at any given time. We run Sinfonia on the cluster in primary-backup replication mode so that each server acts as both a primary node and a backup (to a different memnode). We disable logging in order to maintain all state in memory, though Minuet will work with any of Sinfonia's fault tolerance options at some cost to performance.

To generate a standard repeatable workload, we used the open-source Yahoo Cloud Serving Benchmark (YCSB) [14]. YCSB can generate key-value operations and range scans, and has been used to benchmark a variety of distributed storage systems. In order to ensure that Minuet receives sufficient load, we scale the number of YCSB clients with the size of our experiment. On each server, the YCSB client runs on one processor (six cores) and Minuet on the other, as shown in Figure 9. For the processor running Minuet, we run one memnode on two cores, and one proxy on three cores. The final core is reserved for processing interrupts. The YCSB client on each server sends requests only to the proxy on the local host.

We compare the performance of Minuet against a modern commercial main-memory database, which we will refer to as CDB. We configured this system to emulate a key-value store, and replicated all the data once so that each data item is stored on two servers. This matches the primary-backup replication we used for Minuet. We allocated five cores to CDB, as with Minuet, and configured the system according to its documentation. The YCSB client makes synchronous requests to the CDB process on the same host.

In all our experiments, we configured both Minuet and CDB with 14-byte keys and 8-byte integer values. In Minuet we used the multiversion B-tree described in Sections 3 and 4, with 4kB tree nodes. The CDB schema provides a single table and stored procedures for read, insert and update operations. We implemented our own stored procedure for the YCSB scan operation, which retrieves a given number of consecutive keys (and their values) starting at a given search key. Unless otherwise specified, each data point presented in this section is the average of three trial runs, where the system was pre-loaded with 100 million key-value pairs chosen uniformly at random. Note that we fix the tree size even as we increase the number of machines, and so our results show how throughput scales on a fixed size problem (often called *strong scaling*). Error bars indicate the sample standard deviation of the trials.

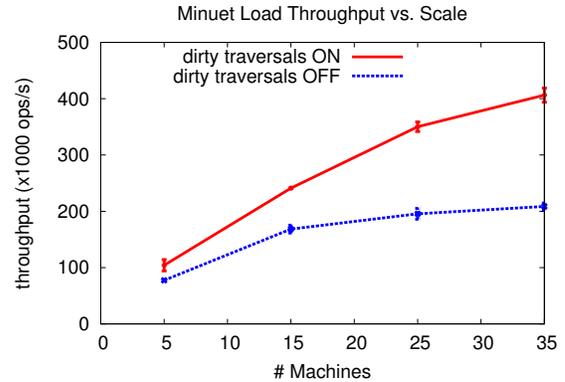

**Figure 10: Minuet load throughput.**

### 6.2 Latency and Throughput

**Dirty traversals.** To evaluate the benefit of dirty traversals, we measure throughput for Minuet when loading uniformly random keys starting with an empty B-tree. We ran the YCSB load phase for exactly 60 seconds and measured the aggregate throughput, as well as mean and 95th %-ile latency. In one run we used dirty traversals and did not replicate any B-tree node sequence numbers. In another run we disabled dirty traversals and replicated inner B-tree node sequence numbers as in Aguilera et al. [5]. The results for various scales are presented in Figure 10.

As shown in the figure, aggregate throughput scales much better with dirty traversals enabled—up to a factor of two better than the version without dirty traversals on 35 hosts. Mean latency (not shown) with dirty traversals is under 2.5 ms at all points, which is up to 1.8 times lower than the version without dirty traversals. However, dirty traversals make relatively little difference on average throughput for uniform single-key operations when the tree is pre-populated, since there is much less contention. We omit the graph for lack of space, and point out that when the workload is skewed, a larger B-tree can experience contention just like the smaller B-tree used in our microbenchmarks. Furthermore, dirty traversals still have the advantage of allowing us to remove the replicated sequence number table. The practical benefits of this are explained in Section 3.

**Latency-throughput trade-off.** In this experiment, we measure the latency per operation of Minuet and CDB. We used a 10-node instance of each system and varied the amount of load offered by YCSB. In order to reach peak throughput, we ran the YCSB client with 64 threads for Minuet and 512 threads for CDB, which yielded the best performance. The latency-throughput curves are shown in Figure 11. For read operations, Minuet mean latency is below 0.4ms at load levels up to 90% of peak throughput, which beats CDB latency by an order of magnitude. (Note that the $y$-axis scale is $10\times$ higher on the plot for CDB than on the plot for Minuet.) Inserts and updates in Minuet take less than 1ms on average for 20% to 80% peak throughput, and their latency is more than an order of magnitude less than in CDB.

Somewhat surprisingly, Minuet has relatively high latency at less than 20% of peak throughput. We speculate that this is due to the network driver tuning the interrupt throttling rate according to the offered load. Measurements of round trip time (using ping) at various loads corroborate this hypothesis.

**Scalability for single-key transactions.** We tested the peak throughput of Minuet against CDB for single-key read, update, and insert workloads. Figure 12 shows aggregate throughput for scales of 5 to 35 hosts. The performance curves show comparable throughput and good scaling for both systems, though Minuet scales slightly more linearly and has slightly faster reads at larger scales. There is also a greater difference in performance between reads and writes (i.e., updates or inserts) in Minuet than in CDB. Reads are up to 50% better than writes in Minuet, but less than 10% faster in CDB.

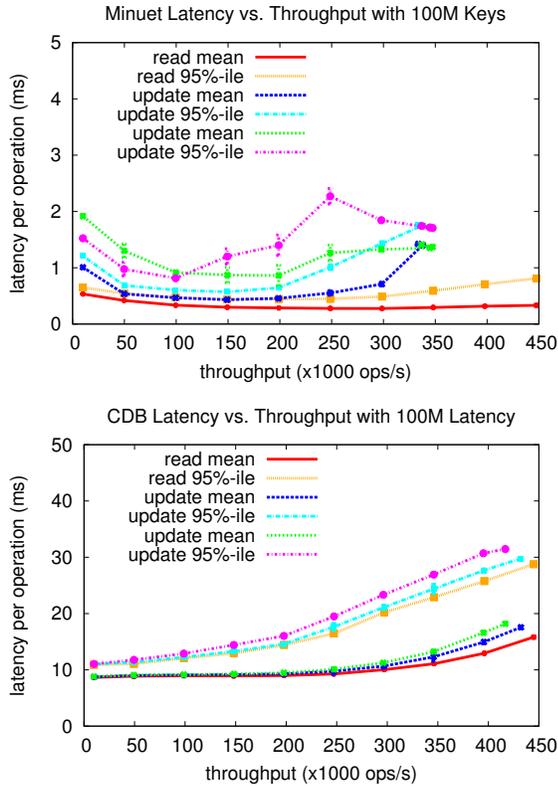

Figure 11: Minuet and CDB latency-throughput trade-off with 15 hosts.

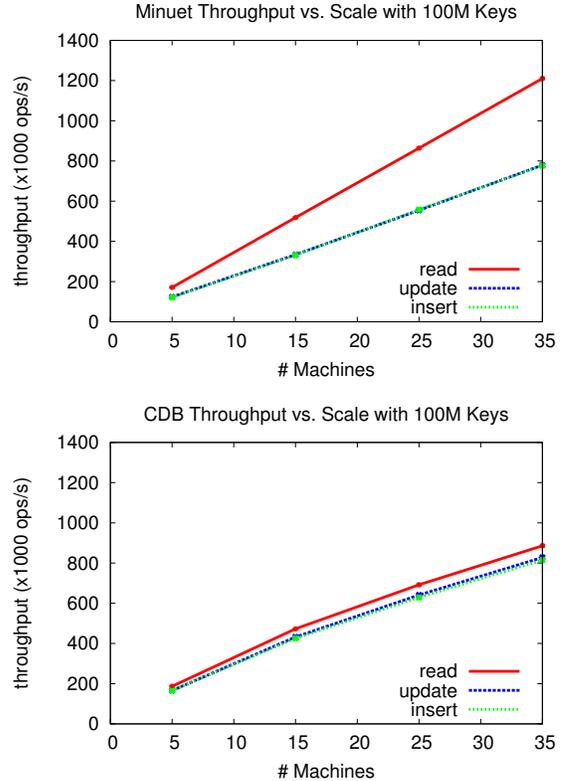

Figure 12: Minuet and CDB throughput scalability for single-key transactions.

**Scalability for multi-index transactions.** We also evaluate the performance of Minuet and CDB for transactions that access more than one index. To that end, we extend the CDB schema with a second table and add stored procedures that atomically read two rows, update two rows, or insert two rows. Each table is hash-partitioned independently and not replicated. Similarly, for Minuet we create two B-trees, and define dynamic transactions that access both B-trees atomically. We extend YCSB to generate operations where the key for each table is drawn uniformly at random. We pre-load each table with 10M keys drawn uniformly at random.

Figure 13 presents the aggregate throughput for dual-key operations in Minuet and our extension of CDB. Minuet scales nearly linearly and at 35 servers performs about 250K dual-key reads, or over 50K dual-key inserts. In contrast, the modified CDB does fewer than 1200 transactions per second and performance drops with scale. This is because each dual-key transaction in CDB engages all servers.

## 6.3 Snapshot Experiments

In Minuet, snapshots are designed to support long running queries concurrently with short transactional operations. In this section, we evaluate how well this technique works and investigate the impact of snapshot creation on the performance of transactions.

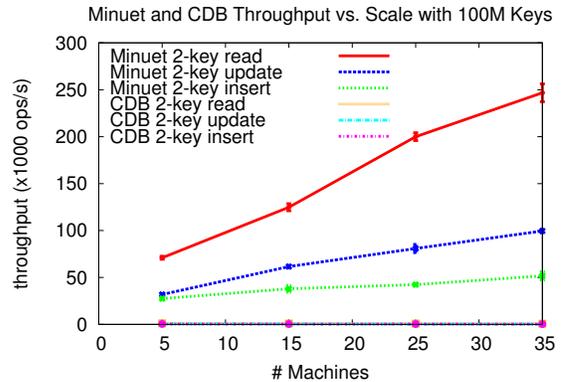

Figure 13: Multi-index throughput.

Figure 14 is a time series showing the impact of a single snapshot on the throughput of a 100% update workload executing on 25 servers. For this experiment, we repeatedly executed a one-second YCSB workload and measured the throughput after each run. We issued a single snapshot request after the fifth run (at 20 s). The results indicate that while snapshot creation is a heavyweight operation, its impact is relatively short-lived, and the throughput returns to its pre-snapshot levels within 20-30s. We observe that snapshots will have a fairly minor effect on the performance of read operations since the latter do not trigger copy-on-write. Thus, Figure 14 demonstrates a worst-case scenario; we would expect less disruption in a workload with a combination of reads and writes instead of 100% updates.

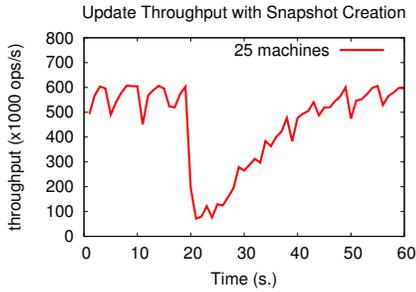
Figure 14: Impact of Snapshot.

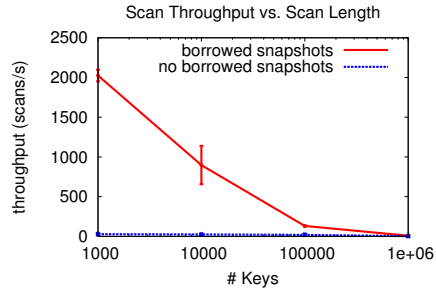
Figure 15: Borrowed snapshots.

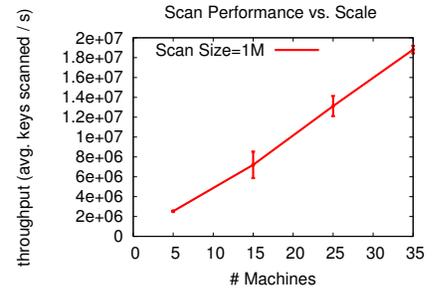
Figure 16: Scalability of scans.

**Range scans with snapshots.** To evaluate how well Minuet supports long-running operations, we issue scans over consecutive keys using YCSB's scan request type. When run in Minuet without snapshots, these long scans may never commit, as any concurrent update or insert within the scan's read set will force the scan to abort. Consequently, we execute scans in Minuet by first creating a snapshot and then running the scan against that snapshot. As noted in Section 4, this ensures that the scans will be strictly serializable and that they will eventually terminate, as snapshots are read only.

To evaluate this technique, we run a 100% update workload as described previously, and concurrently run one additional YCSB client that issues scans from a single thread. By default we set the scan size to 1M keys (10% of the data set).

When the number of scans is large, throughput can suffer considerably due to the use of blocking minitransactions (described in Section 4.1) and contention due to the large number of copy-on-write operations. However, for many applications, it may not be necessary that all long-running operations be strictly serializable; ordinary serializability may suffice. Thus, we also experiment with an additional parameter—the minimum time between snapshots. When this is set to 0, the algorithm behaves as described previously, and a snapshot is taken for every scan, guaranteeing strict serializability. However, when this parameter is set to $k > 0$, a new snapshot will be created at most once every $k$ seconds, and additional scan operations will use the most recent snapshot. In some ways this is a more aggressive form of borrowed snapshots (see Section 4.3), but it does not guarantee strict serializability as the scans may see a consistent view of the database that is out-of-date by as much as $k$ seconds.

Figure 17 shows the impact of long running scans on the update throughput for different values of $k$. The topmost line shows the update throughput without any scans, and is similar to that shown in Figure 12. The remaining lines show the behavior for $k = 0$, $k = 5$, $k = 30$, and $k = 60$. Unsurprisingly, the performance is best when $k$ is large. With $k = 60$, the update throughput is 50-70% of the throughput without scans. As the frequency of snapshots increases (i.e. $k$ gets smaller), snapshot creation becomes a bottleneck, and the update throughput drops significantly. With $k = 0$ (i.e. a snapshot taken at every scan), the throughput is less than 10% of that with no scans.

We also measure the average latency of both scan and update operations when they are running concurrently in the configuration described above. Figure 18 shows the latency of scans with 15 hosts as we vary $k$ from 0 to 60s. We do not show the corresponding update latency due to lack of space, but it follows a simple curve, ranging from approximately 16ms at $k = 0$ to less than 3ms at $k \geq 30$, approaching 2ms at $k = 60$. As expected, the update latency decreases as $k$ increases: since there are fewer snapshots, a smaller fraction of the updates trigger a copy-on-write, and thus they execute more quickly. The behavior of the scan la-

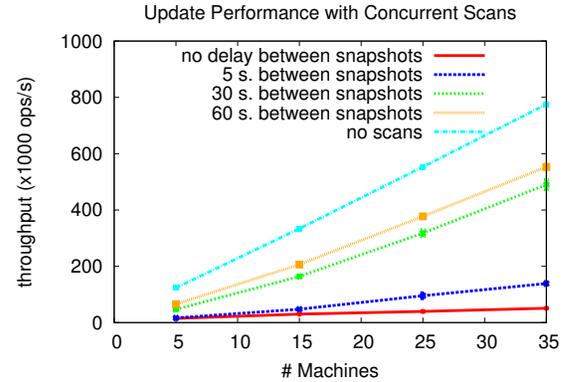
Figure 17: Update throughput with scans.

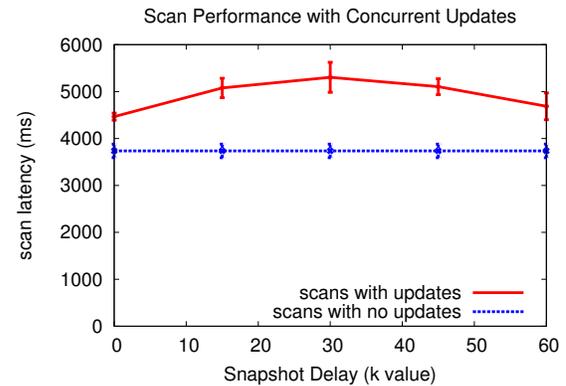
Figure 18: Scan latency with updates.

tency is more complex, which we expect is due to two competing factors. On one hand, as $k$ increases, the fraction of scans that must create a new snapshot decreases, which should decrease the average latency. On the other hand, the update throughput increases with larger $k$, which means that more memnode resources will be used to process updates. This would tend to increase the average scan latency. The combination of these effects leads to the curved shape in Figure 18. Note, however, that the average scan latency with concurrent updates is never more than 1.4 times the scan latency with no concurrent updates, which suggests that snapshots are successful in isolating the scans from the update workload.

**Borrowed snapshots.** While setting $k > 0$ is similar to borrowing snapshots, it does not guarantee strict serializability. To measure the performance of the strictly serializable borrowed snapshot technique described in Section 4.3, we need a workload with some contention between snapshots. For this experiment, we partitioned 15 YCSB clients so that 3 of them run a 100% scan workload, and the 12 execute a 100% update workload. Figure 15 shows the scan throughput for scans of different sizes (note the log scale on the x axis). With relatively short scans (1000 keys), the number of snapshots is very high, and borrowing snapshots improves the throughput by more than an order-of-magnitude. As the scan size increases, however, snapshot creation ceases to be the bottleneck. With scans of 1,000,000 keys, the performance of the two methods is essentially identical.

**Scan scalability.** In our final experiment, we evaluate how well long one-million key scans scale with system size. As snapshot creation will become a bottleneck when $k = 0$, we fix the minimum time between snapshots to $k = 30$s in order to measure the peak throughput with a modest amount of staleness. As in the borrowed snapshot experiment, we partition the YCSB clients so that 80% of them execute updates and the other 20% execute scans. Figure 16 shows the throughput of the scans in keys/s as we scale the total number of servers from 5 to 35. The curve is almost perfectly linear, which suggests that 30s is a large enough snapshot interval to avoid a bottleneck on snapshot creation. We attempted to compare Minuet with CDB on scan performance, but CDB was unable to perform long scans due to internal memory limitations for individual queries. (Our servers had excess DRAM.) Barring this limitation, we expect that CDB would not scale at all because each scan would engage all servers, as in our multi-index operations.

# 7. RELATED WORK

Minuet leverages ideas from many areas of research. In this section we give an overview with a focus on design features and APIs.

**B-trees and multiversioning.** B-trees are ubiquitous and there is a large body of work on optimizing them for specific use cases. Prior work on concurrency control in B-trees focuses mostly on lock-based techniques for centralized implementations [18]. To our knowledge, Lehman and Yao's B-link tree [28] was the first to use fence keys for concurrency control, enabling lock-free lookups. Minuet uses optimistic concurrency control and includes two fence keys per B-tree node, whereas B-link trees require only a high fence. Fence keys have also been used in write-optimized B-trees [17] and for verification of B-tree integrity [19].

A wide variety of multiversion B-trees have been proposed in the literature. Copy-on-write (CoW) methods, like that used by Minuet, are inspired by the path-copying algorithm from Driscoll et al. [16], and have been widely used in file systems [12, 21, 9]. CoW B-trees have good query performance, but updates can be expensive due to path copying [37]. Per-key versioning avoids this problem by storing multiple versions of individual keys in the same B-tree node, but reads may be slower as they may have to traverse multiple versions in each node [27, 29, 7].

Several other proposals for versioned dictionaries are related to Minuet. Log-structured access methods, such as the log structured merge tree (LSM-tree) [33] and the log-structured history data access method (LHAM) [31], are widely used for multiversion data. Recently, Google released LevelDB, a single-node ordered key-value store based on the LSM-tree that supports multi-key updates and read-only snapshots [3]. Unlike Minuet, all of these methods use secondary storage, and are optimized for write (append) operations at the expense of reads. The *stratified B-tree* of Twigg et al. is another recent multiversion data structure [39]. It is cache oblivious and designed to take advantage of fast sequential I/O by storing data using a collection of arrays. Operations on data are extremely fast in the amortized sense, but the worst-case cost of insertions can be high due to splitting and merging of arrays. In contrast to all of these methods, Minuet uses in-memory storage, is read-optimized, and designed for scalability in a shared nothing environment.

Rodeh proposed a multi-version B-tree that supports branching versions, with applications to file systems [36], but relatively few other systems have considered how to make use of branching versions. Landau et al. proposed novel queries for branching versions: *vertical queries* access a version and its ancestors in the version tree, while *horizontal queries* access multiple descendants of the same version [26]. Jiang and Salzberg proposed the BT-tree, which uses per-key versioning to support branching versions with vertical and horizontal queries [23]. None of these projects addressed the problems of concurrency control or scale-out.

**Distributed data structures.** Johnson and Colbrook's data structure and Boxwood [24, 30] are distributed B-trees. Neither provides multiversioning or transactions, and both rely on subtle protocols and locking schemes. Aguilera et al. [5] proposed a transactional B-tree (on which Minuet is based) that is simple conceptually but fails to scale for insertions and lacks multiversioning. Najaran et al. proposed a similar multi-dimensional tree which scales well but is not balanced, so operations run in linear time in the worst case [32].

Many peer-to-peer systems, including file sharing systems, use distributed data structures internally. Most use distributed hash tables, and hence fail to provide efficient ordered traversals. A few support distributed range queries efficiently, including BATON [22], which is based on a balanced binary tree, and P-Ring [15], which uses a ring structure coupled with a B-tree-like splitting algorithm. These projects focus on dealing with the high churn found in peer-to-peer environments, and provide weak consistency.

**Cloud storage and indexing systems.** The number of cloud storage systems has grown rapidly in recent years. The vast majority (e.g., most key-value stores) fail to provide efficient ordered traversals, multiversioning, and consistency guarantees. Google's BigTable [11] (and its open-source variant HBase [2]) and Yahoo's PNUTS [13] use key range partitioning for efficient range queries. These systems are disk-based and support only per-record atomicity. While all of them are multi-versioned, only BigTable and HBase allow users to access specific versions directly. PNUTS also supports materialized views and secondary indexes, but they are maintained asynchronously. In contrast, Minuet supports strictly serializable transactions across multiple keys and indexes.

CouchDB is a distributed data store for JSON data [1]. It uses a B+-tree on each node internally, and supports partial replication and merging, which is somewhat similar to Minuet's branching versions. However, CouchDB provides only eventual consistency and is designed to store documents rather than simple key-value pairs.

Some recent projects propose using overlays like BATON [22] to range partition data for efficient ordered traversals in cloud storage systems. The ec-Store index [40] extends BATON with adaptive replication for load balancing and uses optimistic multiversion concurrency control to provide snapshot isolation. All transactions are serialized at a single node in the system, which limits scalability. (Minuet serializes only snapshot creation operations.) The CG-index [41] also uses BATON, but focuses on providing secondary indexes over data that is already partitioned across a cluster.

**Hybrid OLTP and Analytics.** The HyPer database system uses snapshots to support both OLTP transactions and long-running analytics queries efficiently in a single platform [25]. HyPer is a single-node system that leverages hardware-assisted copy-on-write provided by the virtual memory manager, which enables extremely fast checkpoints for analytics queries. $ES^2$ is a distributed system for OLTP and OLAP, and supports multiversioning for read-only analytics queries [10]. Its index is based on the ec-Store system described earlier. In contrast to these systems, Minuet scales to hundreds of cores and TBs of memory while supporting multiversioning with writable snapshots and branching versions.

**Other storage systems.** VoltDB [4] is a distributed OLTP system based upon H-Store [38]. VoltDB is extremely fast and scalable for workloads that can be partitioned effectively using hashing. In order to reduce synchronization overheads, only one thread can access a given partition. Range queries over partitioned tables are supported but do not scale because each query engages all servers.

Hyder is a write-optimized distributed transaction processing system that leverages flash storage [8]. Hyder marshals a binary-tree index structure into a shared log, and uses copy-on-write to update the index in an append-only manner. In that sense, transactions are executed against a snapshot of the database. Hyder lacks branching versions and does not offer special support for analytics.

## 8. CONCLUSIONS

Minuet allows emerging data-centric businesses to meet strategic imperatives by supporting both transactional and analytic workloads in a single scalable platform. It also offers a powerful feature that enables sophisticated "what-if" analysis, data sharing, archiving, and data protection. Compared to a modern industrial strength main-memory database, Minuet provides superior latency and is competitive on aggregate throughput for simple transactions. Unlike conventional hash-partitioned systems, it also enables scalable analytics queries. Our solution unifies traditionally disparate systems and workloads into a single platform that we hope will enable a whole new class of online use-cases and services.

**Acknowledgments.** Sincere thanks to Harumi Kuno and Goetz Graefe for stimulating discussions about B-trees; to Stavros Harizopoulos for his helpful suggestions regarding analytics and snapshotting; and to John Byrne, Nathan Binkert, Eric Anderson, Joe Tucek and Terence Kelly for their help with optimizing our implementation.

## 9. REFERENCES

[1] CouchDB. http://couchdb.apache.org/.
[2] HBase. http://hbase.apache.org/.
[3] LevelDB. http://code.google.com/p/leveldb/.
[4] VoltDB. http://voltdb.com/.
[5] M. K. Aguilera, W. Golab, and M. A. Shah. A practical scalable distributed B-tree. *PVLDB*, 1(1):598–609, 2008.
[6] M. K. Aguilera, A. Merchant, M. A. Shah, A. C. Veitch, and C. T. Karamanolis. Sinfonia: A new paradigm for building scalable distributed systems. *ACM Trans. Comput. Syst.*, 27(3):5:1–5:48, 2009.
[7] B. Becker, S. Gschwind, T. Ohler, B. Seeger, and P. Widmayer. An asymptotically optimal multiversion B-tree. *VLDB J.*, 5(4):264–275, 1996.
[8] P. A. Bernstein, C. W. Reid, and S. Das. Hyder - a transactional record manager for shared flash. In *Proc. CIDR*, pages 9–20, 2011.
[9] J. Bonwick and M. Ahrens. The Zettabyte file system. 2008.
[10] Y. Cao, C. Chen, F. Guo, D. Jiang, Y. Lin, B. C. Ooi, H. T. Vo, S. Wu, and Q. Xu. ES$^2$: A cloud data storage system for supporting both OLTP and OLAP. In *Proc. ICDE*, pages 291–302, 2011.
[11] F. Chang, J. Dean, S. Ghemawat, W. C. Hsieh, D. A. Wallach, M. Burrows, T. Chandra, A. Fikes, and R. E. Gruber. Bigtable: A distributed storage system for structured data. *ACM Trans. Comput. Syst.*, 26(2):4:1–4:26, 2008.
[12] S. Chutani, O. T. Anderson, M. L. Kazar, B. W. Leverett, W. A. Mason, and R. N. Sidebotham. The Episode file system. In *Proc. USENIX ATC*, pages 43–60, 1992.
[13] B. F. Cooper, R. Ramakrishnan, U. Srivastava, A. Silberstein, P. Bohannon, H.-A. Jacobsen, N. Puz, D. Weaver, and R. Yerneni. Pnuts: Yahoo!'s hosted data serving platform. *PVLDB*, 1(2):1277–1288, 2008.
[14] B. F. Cooper, A. Silberstein, E. Tam, R. Ramakrishnan, and R. Sears. Benchmarking cloud serving systems with YCSB. In *Proc. SoCC*, pages 143–154, New York, NY, USA, 2010. ACM.
[15] A. Crainiceanu, P. Linga, A. Machanavajjhala, J. Gehrke, and J. Shanmugasundaram. P-ring: an efficient and robust p2p range index structure. In *Proc. SIGMOD*, pages 223–234, 2007.
[16] J. R. Driscoll, N. Sarnak, D. D. Sleator, and R. E. Tarjan. Making data structures persistent. *J. Comput. Syst. Sci.*, 38(1):86–124, 1989.
[17] G. Graefe. Write-optimized B-trees. In *Proc. VLDB*, pages 672–683, 2004.
[18] G. Graefe. A survey of B-tree locking techniques. *ACM Trans. Database Syst.*, 35(3):16:1–16:26, July 2010.
[19] G. Graefe and R. Stonecipher. Efficient verification of B-tree integrity. In *Proc. BTW*, pages 27–46, 2009.
[20] T. Härder. Observations on optimistic concurrency control schemes. *Inf. Syst.*, 9(2):111–120, June 1984.
[21] D. Hitz, J. Lau, and M. A. Malcolm. File system design for an NFS file server appliance. In *USENIX Winter*, pages 235–246, 1994.
[22] H. V. Jagadish, B. C. Ooi, and Q. H. Vu. BATON: A balanced tree structure for peer-to-peer networks. In *Proc. VLDB*, pages 661–672, 2005.
[23] L. Jiang, B. Salzberg, D. B. Lomet, and M. B. García. The BT-tree: A branched and temporal access method. In *Proc. VLDB*, pages 451–460, 2000.
[24] T. Johnson and A. Colbrook. A distributed data-balanced dictionary based on the B-link tree. In *Proc. IPPS*, pages 319–324, 1992.
[25] A. Kemper and T. Neumann. Hyper: A hybrid OLTP&OLAP main memory database system based on virtual memory snapshots. In *Proc. ICDE*, pages 195–206, 2011.
[26] G. M. Landau, J. P. Schmidt, and V. J. Tsotras. Historical queries along multiple lines of time evolution. *VLDB J.*, 4(4):703–726, 1995.
[27] S. Lanka and E. Mays. Fully persistent B+-trees. In *Proc. SIGMOD Conference*, pages 426–435, 1991.
[28] P. L. Lehman and S. B. Yao. Efficient locking for concurrent operations on B-trees. *ACM Trans. Database Syst.*, 6(4):650–670, December 1981.
[29] D. B. Lomet and B. Salzberg. Access methods for multiversion data. In *Proc. SIGMOD*, pages 315–324, 1989.
[30] J. MacCormick, N. Murphy, M. Najork, C. Thekkath, and L. Zhou. Boxwood: Abstractions as the foundation for storage infrastructure. In *Proc. OSDI*, pages 105–120, Dec. 2004.
[31] P. Muth, P. E. O'Neil, A. Pick, and G. Weikum. The LHAM log-structured history data access method. *VLDB J.*, 8(3-4):199–221, 2000.
[32] M. T. Najaran, P. Wijesekera, A. Warfield, and N. C. Hutchinson. Distributed indexing and locking : In search of scalable consistency. In *Proc. LADIS*, 2011.
[33] P. E. O'Neil, E. Cheng, D. Gawlick, and E. J. O'Neil. The log-structured merge-tree (LSM-tree). *Acta Inf.*, 33(4):351–385, 1996.
[34] C. H. Papadimitriou. The serializability of concurrent database updates. *J. ACM*, 26(4):631–653, 1979.
[35] D. Reed. Naming and synchronization in a decentralized computer system, 1978.
[36] O. Rodeh. B-trees, shadowing, and clones. *Trans. Storage*, 3(4):2:1–2:27, 2008.
[37] C. A. N. Soules, G. R. Goodson, J. D. Strunk, and G. R. Ganger. Metadata efficiency in versioning file systems. In *Proc. FAST*, pages 43–58, 2003.
[38] M. Stonebraker, S. Madden, D. J. Abadi, S. Harizopoulos, N. Hachem, and P. Helland. The end of an architectural era (it's time for a complete rewrite). In *Proc. VLDB*, pages 1150–1160, 2007.
[39] A. Twigg, A. Byde, G. Milos, T. D. Moreton, J. Wilkes, and T. Wilkie. Stratified B-trees and versioning dictionaries. *CoRR*, abs/1103.4282, 2011.
[40] H. T. Vo, C. Chen, and B. C. Ooi. Towards elastic transactional cloud storage with range query support. *PVLDB*, 3(1):506–517, 2010.
[41] S. Wu, D. Jiang, B. C. Ooi, and K.-L. Wu. Efficient B-tree based indexing for cloud data processing. *PVLDB*, 3(1):1207–1218, 2010.